\documentclass[sigconf, natbib=false,
]{acmart}

\pdfoutput=1 

\RequirePackage[
  datamodel=acmdatamodel,
  style=acmnumeric, 
  sortcites=true, isbn=false]{biblatex}

\usepackage{csquotes}
\usepackage{graphicx}

\makeatletter
\def\maxwidth{\ifdim\Gin@nat@width>\linewidth\linewidth
\else\Gin@nat@width\fi}
\makeatother
\let\Oldincludegraphics\includegraphics
\renewcommand{\includegraphics}[1]{\Oldincludegraphics[width=\maxwidth]{#1}}

\setcopyright{none}
\renewcommand\footnotetextcopyrightpermission[1]{}
\copyrightyear{2023}
\acmYear{2023}
\acmConference[in2writing]{The Second Workshop on Intelligent and Interactive Writing Assistants}{April 23, 2023}{Hamburg, Germany}
\acmBooktitle{The Second Workshop on Intelligent and Interactive Writing Assistants, April 23, 2023, Hamburg, Germany}

\ccsdesc[500]{Human-centered computing~Text input}
\ccsdesc[500]{Human-centered computing~User interface design}
\ccsdesc[300]{Human-centered computing~Graphical user interfaces}
\ccsdesc[300]{Human-centered computing~Hypertext / hypermedia}
\ccsdesc[500]{Applied computing~Document preparation}

\keywords{writing technology, natural language processing, intelligent writing tools, interactive editing}

	\author{Cerstin Mahlow}
			\orcid{https://orcid.org/0000-0003-0215-5551}
		\affiliation{
					\institution{School of Applied Linguistics, Zurich University of
Applied Sciences}
									\city{Winterthur}
							\country{Switzerland}
			}
	\email{cerstin.mahlow@zhaw.ch}

\newcommand{\tightlist}{\relax} 

\begin{document}

	\title{Writing Tools: Looking Back to Look Ahead}

\begin{abstract}
Research on writing tools started with the increased availability of
computers in the 1970s. After a first phase addressing the needs of
programmers and data scientists, research in the late 1980s started to
focus on writing-specific needs. Several projects aimed at supporting
writers and letting them concentrate on the creative aspects of writing
by having the writing tool take care of the mundane aspects using NLP
techniques. Due to technical limitations at that time the projects
failed and research in this area stopped. However, today's computing
power and NLP resources make the ideas from these projects technically
feasible; in fact, we see projects explicitly continuing from where
abandoned projects stopped, and we see new applications integrating NLP
resources without making references to those old projects. To design
intelligent writing assistants with the possibilities offered by today's
technology, we should re-examine the goals and lessons learned from
previous projects to define the important dimensions to be considered.
\end{abstract}

\renewcommand{\shortauthors}{}

\maketitle

\hypertarget{where-we-come-from}{%
\section{Where we come from}\label{where-we-come-from}}

The term \enquote{word processing} first appeared in the 1960s,
referring to a combination of hardware and software \autocites[for the
history of word processing see,][]{eisenberg1992,haigh2006}. The first
writing tools for microcomputers, such as Electric Pencil or Easy
Writer, were developed by hobby programmers; they were soon displaced by
commercial products \autocite{bergin2006a}. How people actually used
these new tools was not obvious. \textcite{rosson1983} and
\textcite{whiteside1982} therefore asked \enquote{How do people really
use text editors?} (title of \autocite{whiteside1982}) in the early
1980s to gain insights for the development of future editors. But even
ten years later, after the failure of a large project on writing
support, \textcite{holt1992} had to admit:

\begin{quote}
In seems clear {[}\ldots{]} that in order to produce computer based
tools to support writers and the writing process we must increase our
knowledge of how writers conduct their craft. An increased understanding
of writer's requirements and the task involved in writing will form the
basis of the next generation of writing tools. \autocite[X]{holt1992}
\end{quote}

Projects like RUSKIN \autocite{williams1989,williams1990}, Writer's
Assistant \autocite{sharples1990,sharples1989}, Intelligent Workstation
\autocite{kempen1986}, and Editor's Assistant
\autocite{dale1989,dale1996} did not result in marketable products. They
aimed to improve (post-)editing and revision based on linguistic
principles using linguistic resources. The design and development,
however, did not take into account the real needs of users. Natural
language processing (NLP) resources were not yet mature enough to be
used in real applications. The computing power of PCs at the time was
insufficient for real-time analysis---Editor's Assistant relied on
constant full syntactic parsing of the growing text---and generation and
the resulting tools were thus too limited for practical use.

For RUSKIN, a post-editing support tool, \textcite{williams1990a} notes
that at the time, most writing software was rather ad hoc
implementations of ideas with poor user interfaces incorporating
checkers that provided inadequate or even erroneous results
\autocite[3]{williams1990a}. The project aimed to overcome the poor
quality of checkers by incorporating automatic syntax analysis---the
probabilistic parser only delivered results with a correctness of around
75\%, though \autocite[117]{williams1993}. Furthermore, the supposed
\enquote{authors' needs} were based solely on the researchers'
intuitions. Only at the end of the project they realized that real
authors actually would have preferred support \emph{during} writing, not
in a separate post-editing phase, as \enquote{the concept of postwriting
software implies a linear model of the human writing process which is at
best simplistic and at worst may be completely misleading}
\autocite[6]{williams1989}. The first models of writing as a complex and
non-linear process had already been published at this point, e.g., by
\textcite{flower1981} and \textcite{scardamalia1983}. Writer's Assistant
did take into account writing research, but lacked an
\emph{implementable} writing model \autocite[26]{sharples1989}. This is
still an open issue in writing research today, but only rarely addressed
\autocites[e.g.,][]{hayes2001,hayes2012b}. The prototype of Writer's
Assistant \autocite{sharples1990} focused on authors' activities and
aimed to be more than just the next word processor: \enquote{The
Writer's Assistant is a computer-based \emph{cognitive support} system
for people who create complex documents as part of their professional
live} \autocite[p.~22, emphasis in original]{sharples1989}. Writer's
Assistant was supposed to be a combination of word processor, idea
processor, and outliner/structure editor to provide authors with
different views on different aspects on the text under development
\autocite[22]{sharples1989}. But at that time, it could not be
implemented and remained a thought model.

Serious research on writing tools stopped in the late 1980s, once
corporate customers---including universities---had decided what to
purchase
\autocites[see][]{daiute1983,taylor1987,hawisher1988,ross1991,vernon2000}
and Microsoft Word had achieved monopoly status in the consumer market
\autocites[see][]{eisenberg1992,bergin2006b,wohl2006}. The main reason
was that MS Word was bundled with many PCs. From a customer's
perspective, everything was ready: one could just turn on the new
computer and start writing with MS Word. Purchasing and installing
another word processor such as WordPerfect would have required
purchasing another license and installing another program. This ubiquity
had several effects: writers became accustomed to the appearance,
features, and affordances of MS Word; the format of text files produced
with MS Word became the default file format expected and demanded for
submissions of academic theses and the like, for interchange between
writers when writing collaboratively, and for further processing in
publishing houses.

The first effect led to the general assumption that any other new
writing facility, e.g.~in the first learning management systems that
appeared in the early 2000s, should be designed to resemble the look and
feel of MS Word and include its main features to provide a familiar user
experience. This also applies to the first versions of Google's
web-based word processor Google Docs, which became available in a beta
version in early 2006. Experimental projects such as the British
Telecom-funded Editor's Assistant had no progress in sight in the 1990s
to overcome technological obstacles (computing power, quality of NLP)
that would justify further investment.

The integration of NLP technology into word processors beyond checkers
for spelling and grammar has been a research topic since the 1980s
\autocites[e.g.,][]{kempen1986,kempen1992}, but did not result in
commercial products either. To overcome the challenges for parsers
arising from what \textcite{vandevanter1995} calls \enquote{the three
I's}: \emph{ill-formedness}, \emph{incompleteness}, and
\emph{inconsistency} of sentences during writing, experimental word
processors attempted to incorporate \emph{syntax orientation} as derived
from \emph{syntax-oriented} text editors such as EMILY
\autocite{hansen1971}, Cornell \autocite{teitelbaum1981}, PEN
\autocite{allen1981}, JANUS \autocite{chamberlin1981}, PARSE
\autocite{chusho1983}, Mentor \autocite{gouge1983,lang1986}, PAN
\autocite{ballance1992,vandevanter1992}, or CodeProcessor
\autocite{vandevanter2000}. These editors handled documents as tree
structures and were implementations of programming principles like
\emph{stepwise refinement} and \emph{structured programming}
\autocite{dijkstra1972a,wirth1971}. However, similar to programmers,
writers objected to always produce complete, well-formed sentences, as
this was not compatible with their writing habits. It also does not
reflect the writing process as has been observed in various studies:
authors often start revising a sentence \emph{before} a complete first
version of this sentence is finished
\autocites[see][]{olive2009,wengelin2009,leijten2010a}.
\textcite{dale1997} predicted in 1997:

\begin{quote}
The major developments in the next five to ten years are likely to be of
an augmentative nature, with increasingly sophisticated systems that
have people and machines doing what they each do best. The key here is
to add intelligence and sophistication to provide \emph{language
sensitivity}, enabling the software to see a text not just as a sequence
of characters, but as words and sentences combined in particular
structures for particular semantic and pragmatic effect.
\autocite[p.~235, emphasis in original]{dale1997}
\end{quote}

No such systems were available for the general public in the 2000s,
though. At that time, \textcite{mahlow2008a} proposed language-aware
functionality, but only developed a proof of concept as extension to
Emacs \autocite{mahlow2009e}. In writing research, the influence of the
writing tool and medium are only occasionally acknowledged
\autocite{sharples1996c,calonne2006,mahlow2014a}; the field concentrates
on cognitive aspects and writing strategies.

\hypertarget{where-we-are-now}{%
\section{Where we are now}\label{where-we-are-now}}

Starting in the 2010s, the emphasis on writing experience,
personalization of tools, and the growing diversity of input devices
(and methods) and displays prompted the development of \enquote{new
writing tools.} Their functionalities are often working implementations
of methods and concepts originally described and developed in the 1960s
and 1970s that used to be considered failures---but had actually only
failed due to the limitations of computers at that time. Only now we see
the inverse development, back to ideas and applications of the 1960s,
when projects like NLS (oN-Line System) \autocite{engelbart1962} where
\enquote{pushed aside in favor of computer systems more oriented toward
print practices} \autocite{ittersum2008}. NLS already combined
functionality to write text, messages similar to what later became known
as e-mail, and \enquote{computer conferencing} for allowing
collaborative simultaneous editing of documents
\autocite{callender1982}. There was no fixed final document
format---e.g., a printed page---the focus was on facilitating online
text production by implementing

\begin{quote}
text editing capabilities of later word processors, including word wrap,
search and replace, and scrolling, and the use of a mouse to select text
to be cut and pasted between documents. Indeed Engelbart's system was
much more complex than most of subsequent word processing systems
\autocite[21]{haigh2006}
\end{quote}

One type of functionality that \textcite{mahlow2008a} suggest are
information functions that use NLP techniques to highlight certain
aspects of the evolving text, commonly referred to as \enquote{syntax
highlighting.} Since 2013, iA Writer has offered such a feature in a
commercial product to specifically highlight nouns, verbs, adjectives,
etc., advertising it as \enquote{using parts of speech to improve your
writing} and explicitly stating that writers deserve the same
professional support as programmers.\footnote{\url{https://ia.net/writer/support/writing-tips/parts-of-speech}}
The use of NLP has been feasible for some time now, both in terms of
quality and the computing power required.

\textcite{williams1992} stated that professional writers, including
academics and journalists, seemed to be satisfied with the tools
available in the early 1990s. They had adapted to these tools and did
not seem to be aware of other options. In the early 2000s, only writers
who had used WordPerfect or other word processors \enquote{back in the
days} sometimes complained about missing functionality in MS Word.

Today, users are willing to try out new interfaces and new writing
experiences. The implementation of applications with appealing user
interfaces is easier than ever: current programming languages and
toolkits allow for fast development and roll-out of responsive
applications. At the same time, the assumption that any writing tool
must resemble MS Word is fading, which is also driven by developments in
creating, sharing, and accessing documents beyond the paper-based
structure \autocite{mahlow2022b,mahlow2022c}.

For some time now, new writing applications as stand-alone tools or
integrated into other services---e.g., learning management systems or
blogging software---are being developed. As for the first wave of
writing tools, we also see the adoption of tools originally intended for
writing code now for writing all kinds of texts. The shift of academic
writing to include dynamic aspects of \enquote{text,} e.g., code
(snippets), data plots, and other visualizations clearly supports the
use of these affordances.

\hypertarget{where-we-should-be-going}{%
\section{Where we should be going}\label{where-we-should-be-going}}

The failed projects from the late 1980s addressed issues that can be
considered general considerations for the design and implementation of
writing technology:

\begin{enumerate}
\def\labelenumi{\arabic{enumi}.}
\tightlist
\item
  user-friendly interfaces, carefully designed functionality instead of
  ad-hoc hacks \autocite{williams1990a};
\item
  support not only for writing, but also for teaching and learning how
  to write \autocite{williams1989}, including sophisticated feedback on
  various levels to stimulate reflection on the writing and decisions
  for revising and editing \autocite{holt1990};
\item
  help for interpreting system messages and feedback
  \autocite{mcgowan1992};
\item
  easy extension of features based on user needs \autocite{holt1990};
\item
  real interaction with the system that enables writers to stay in
  control of edits \autocite{dale1990a};
\item
  application of NLP technology users can trust \autocite{dale1996};
\item
  various views on the evolving text (rhetoric, linguistic,
  typographical, graphemic) to stimulate creativity
  \autocite{kempen1986}.
\end{enumerate}

Most of these requirements are generic requirements for software
development and emphasize the need for input from real users, both for
functionality and for the user interface. Strong collaboration between
designers/developers of writing tools and writing researchers modeling
human writing processes at multiple levels (e.g., the cognitive or the
linguistic level) should be established.

We are already seeing experimental applications that use recent
technological possibilities to finally approach writing in ways that
previous experiments could not realize: One such example is
\href{https://twitter.com/tilioapp}{Tilio}, which tried to implement the
ideas proposed by \textcite{sharples1999} by understanding writing as
design and incorporating aspects and techniques today known as
\enquote{design thinking.} While this endeavor was halted by the
COVID-19 situation in 2020, the technical feasibility was demonstrated
in an alpha version, so we may see another attempt later. Similarly, the
combination of different features and services in an application like
Scrivener for seamless integration of idea creation, management of
sources and references, connection to data tracing, and communication
channels (chat and messaging) can be seen as a functional implementation
of \textcite{engelbart1962}'s ideas, even if the developers do not
explicitly refer to it.

In the late 1990s, projects like Intelligent Workstation, intended as an
instance of the \enquote{fifth generation of text-editing programs}
\autocite{kempen1986}, and Integrated Language Tools for Writing and
Document Handling from KTH Stockholm suffered from insufficient NLP
resources. Today's NLP tools make it worth to reconsider the underlying
ideas of those projects. They are also of interest for document creation
processes, as they already abstract from the print-oriented document,
which is in line with current developments: the creation of texts for
documents that can be rendered according to need and display device.

The Web generally allows for \emph{dynamic} documents with respect to
form and content. Linking of documents as hypertext challenges authors
during writing but can be supported using recommender functionality
based on artificial intelligence (AI). The understanding of
\enquote{text} changed at the turn of the century to include
\enquote{interactive, hypertextual documents---many of which reside on
the Internet---{[}which{]} use color, sound, images, video, words, and
icons to express their messages} \autocite[282]{bazerman2001}. This
clearly requires tools that allow writers to create and edit such
documents; here again, writers could be supported by powerful AI-based
components.

Taking into account that communication takes place on various channels
with specific and complex formats emphasizes the need for structure
within texts. This allows the display of the content/text according to
features of devices and tailored to the needs of readers. Writing in
these scenarios used to be challenging and required knowledge of
specific markup for rendering. Abandoning WYSIWYG and its focus on
printed paper documents, together with the development of truly
augmented and responsive writing tools based on generative AI could
actually free writers to \enquote{fully embracing the new opportunities
offered by digital media} \autocite{atzenbeck2021}.

\textcite{dale2021} analyze the \enquote{automated writing assistance
landscape in 2021.} GPT-3 was already available at that time and was
integrated into several tools aimed at supporting writers as co-authors.
These applications addressed specific genres like blog posts and poetry,
and specific writing tasks like expanding, rewriting, and shortening
texts \autocite{dale2021}. Some months later, they were included as
writing aids into experimental editors
\autocites[e.g.,][]{dang2022,yuan2022}. However, they were not widely
used and did not trigger the same discussions that we see now. We also
see reimplementations of popular applications with integrated access to
LLMs. One such example is \href{https://lex.page}{Lex}, intended as a
\enquote{Google Docs style editor} \autocite{justin2022}. It has access
to GPT-3 and GPT-4 so that writers can invoke the language model to
produce plausible continuations of the text, taking into account
everything before the current cursor position, and to rewrite and
summarize paragraphs. However, research from a writing research
perspective on how humans and AI-based language models produce text
through \emph{co-creation} is still pending at this point.

In contrast to Dale's 1997 prediction of augmented language-sensitive
editors within 10 years, Dale's 2021 prediction seems feasible and even
close to reality, given the current pace of development in both machine
learning-based NLP and writing application implementations:

\begin{quote}
But the big shift is the transition from tools that help with editing to
tools that help with authoring. It's conceivable that, in 5 years' time,
no automated writing assistance tool will be considered complete without
a functionality that finishes your sentences, writes you out of tight
corners and fills in background paragraphs while you doze at the
keyboard. And given Microsoft's exclusive licencing deal with OpenAI in
regard to GPT-3, it won't be a surprise if, before too long, we see some
of these capabilities as yet another item on Microsoft Word's ribbon
menu. \autocite[518]{dale2021}
\end{quote}

Note that this prediction does not include the part of
language-sensitive or language-aware functionality supporting authors
during production and revision for semantic and pragmatic aspects.

\hypertarget{conclusion}{%
\section{Conclusion}\label{conclusion}}

To design and implement writing tools effectively and efficiently, HCI
researchers must work closely with writing researchers to both foster
the development of operationalizable writing models and base the
implementation of writing tools on the latest insights into writing
processes. Many ideas for designing writing tools that actually address
the needs of writers can be gleaned from earlier projects by exploring
the technical feasibility of the underlying concepts. In this way, the
development of writing tools would finally respond to the demands and
predictions made in the 1990s by \textcite{holt1992} and
\textcite{dale1997}.

\printbibliography

\end{document}